\documentclass[a4paper,11pt]{article}%
\usepackage{geometry}
\usepackage{amsmath}
\usepackage{amsfonts}
\usepackage{amssymb}
\usepackage{graphicx}
\usepackage{indentfirst}
\usepackage{bbold}
\usepackage[small,bf]{caption}
\usepackage{slashed}
\usepackage{xcolor}
\providecommand{\U}[1]{\protect\rule{.1in}{.1in}}

\DeclareMathOperator{\Tr}{Tr}

\newcommand{\p}{\partial}
\newcommand{\tr}{\ensuremath{\mathrm{Tr}}}
\newcommand{\e}{\ensuremath{\mathrm{e}}}

\begin{document}

\date{}
\title{\textbf{Regimes of $3D$ Yang--Mills theory in the presence of a constant vector background}}
\author{
\textbf{D.~R.~Granado}$^{a,b}$\thanks{diegorochagranado@duytan.edu.vn}\,\,,
\textbf{A. J. G. Carvalho$^{c}$\thanks{gomescarvalhoantoniojose@gmail.com}}\,\,
\textbf{A. Yu. Petrov}$^{c}$\thanks{petrov@fisica.ufpb.br}\,\,,
\textbf{David Vercauteren}$^{a,b}$\thanks{vercauterendavid@duytan.edu.vn}\\[2mm]
{\small \textnormal{$^{a}$  \it Institute of Research and Development,}} \\ 
{\small \textnormal{ \it Duy T{a}n University, D{a} Nang 550000, Vietnam}\normalsize} \\ 
{\small \textnormal{$^{b}$  \it Faculty of Natural Sciences,}} \\ 
{\small \textnormal{ \it Duy T{a}n University, D{a} Nang 550000, Vietnam}\normalsize} \\ 
{\small \textnormal{$^{c}$   \it Departamento de F\'{i}sica, Universidade Federal da Para\'{i}ba,}}
 \\ \small \textnormal{ \it Caixa Postal 5008, 58051-970, Jo\~{a}o Pessoa, Para\'{i}ba, Brazil}\normalsize}
\maketitle

\begin{abstract}
In this paper, we take into account the Gribov copies present in 3D Yang–Mills–Higgs theory
with a constant vector background whose presence breaks the Lorentz symmetry. The constant vector background is introduced within the
non-Abelian aether term. Here, we show that this term arises as a one-loop correction. The influence
of the aether coupling constant on the system is treated afterwards. As a result, we find that for some values
of it the theory can be driven from a nonperturbative regime to a perturbative one. In this paper, we
work with the Higgs field in the fundamental representation and in the Landau gauge.
\end{abstract}

\section{Introduction}
The possibility of Lorentz symmetry violation has recently attracted a lot of attention from a variety of viewpoints. It was proposed for the first time in the context of QED by Carroll, Field and Jackiw (CFJ) in the 90s: in \cite{CFJ} they suggested a consistent Lorentz-violating (LV) extension of QED involving a constant axial vector $b_\mu$. This $b_\mu$ vector induces a privileged space-time direction and therefore breaks the Lorentz symmetry. Soon after that, a variety of LV extensions of the standard model were {proposed} \cite{ColKost1,ColKost2}, and many nontrivial issues have come under investigation. Among these, one can mention birefringence and rotation of the polarization of an electromagnetic wave in vacuum (for example \cite{Ja}) shown to take place in various LV extensions of QED (for example \cite{MP,Casana}), ambiguities in the quantum corrections (for example \cite{Ja1}), and perturbative generation of new LV terms (see for example \cite{CFJ}). Plenty of experimental measurements hunting for Lorentz symmetry breaking have been carried out (see for example \cite{Kostelecky:2008ts} and references therein). The renormalizability of minimal LV QED was furthermore discussed in \cite{Santos:2016bqc}. 

Given the importance of the Higgs mechanism in the Standard Model, Lorentz breaking in the $U(1)$ gauge-Higgs system has also enjoyed considerable interest in the last twenty years, with many different scenarios and aspects having come under scrutiny \cite{BaetaScarpelli:2003yd,Anderson:2004qi,Belich:2004pi,Altschul:2012ig,Brito:2013npa,Casana:2014dfa}.

The works listed above considered Lorentz symmetry breaking in the context of QED. At the same time, further development of these studies implied in a formulation of the Standard Model Extension (SME) \cite{ColKost1,ColKost2} involving LV extensions both for Abelian and non-Abelian sectors of the standard model. In particular, non-Abelian generalization of the CFJ term and other relevant terms have been proposed. Further, a list of these terms has been extended by adding higher-derivative ones, see a detailed discussion in \cite{KosLi}. As a result, one can study various impacts of these terms, including non-perturbative ones.

 The non-Abelian {generalization of the} Carroll--Field--Jackiw (CFJ) term can be generated perturbatively (see \cite{ourYM} for more details) and some consequences of adding this term have been discussed \cite{Santos:2016dcw,Santos:2016uds,Santos:2014lfa,ColMac}. The renormalizability of some non-Abelian systems involving additive terms has been explored as well \cite{Santos:2016dcw,Santos:2016uds,Santos:2014lfa}. Recently, the authors of \cite{Granado:2017xjs} studied the path integral quantization of the YM+CFJ system.

When dealing with non-Abelian gauge theories, however, correct treating the nonpertubative regime remains one of the greatest challenges in quantum field theory. It is well known that the perturbative formalism fails for non-Abelian gauge theories at low energy and in the absence of the Higgs mechanism (or with only a small Higgs vacuum expectation value), since the coupling constant becomes strong. To get reliable results in the infrared (IR) limit in the continuum formulation, non-perturbative methods are required. The papers \cite{Roberts:1994dr,Alkofer:2000wg,Cucchieri:2007md,Dudal:2007cw,Dudal:2008sp,Aguilar:2008xm,Comitini:2017zfp} give a small selection of such methods and the results obtained with them. A number of studies over the past decade have focused on the gluon, quark and ghost propagators in the infrared region, where color degrees of freedom are confined. Although these objects are unphysical by themselves (they are gauge dependent), they are nevertheless the basic building blocks, next to the interaction vertices, entering gauge invariant objects directly linked to physically relevant quantities such as the spectrum, decay constants, and critical exponents and temperatures.

Notice that the continuum formulation requires gauge fixing, which, in non-Abelian theories, is much less trivial than in QED \cite{Gribov:1977wm,Singer:1978dk}. At the end of the 70's, Gribov showed that the Faddeed--Popov gauge fixing procedure was not enough to fix all the gauge copies in Yang--Mills theory \cite{Gribov:1977wm}. These extra gauge copies are called ``Gribov copies'', proved by Gribov to influence the system only in the low-energy regime of the theory. The new gauge fixing procedure proposed by Gribov imposes a restriction in the functional integral and leads to a direct modification of the gluon propagator: the gluon propagator exhibits complex poles instead of real ones. The presence of these complex poles suggests that the degrees of freedom have become unphysical or are confined \cite{Sobreiro:2005ec,Vandersickel:2012tz}. Since Gribov's original paper \cite{Gribov:1977wm}, the subject has been developed even more \cite{Zwanziger:1988jt,Zwanziger:1989mf,Zwanziger:1992qr,Dudal:2011gd,Cucchieri:2007rg,Cucchieri:2011ig,Dudal:2010cd,Fukushima:2013xsa,Canfora:2013kma} (see also the reviews \cite{Sobreiro:2005ec,Vandersickel:2012tz}). In \cite{Capri:2012ah,Capri:2013oja} the transition between a confining and a Higgs regime using the Gribov restriction was studied. Similar analyses were made for $\mathcal{N}=1$ super-Yang--Mills \cite{Capri:2014xea}, Yang--Mills--Chern--Simons in 3 spacetime dimensions \cite{Canfora:2013zza}, and the Yang--Mills-aether system \cite{Granado:2018bcp}. All these results enable us to state that the issue of the Gribov copies  captures nontrivial aspects of the non-perturbative dynamics of Yang--Mills theories.

One setting that has not yet been investigated is that of a Lorentz-breaking extension of a non-Abelian Yang--Mills--Higgs theory. In order to rectify this, in this paper we consider $SU(N)$ Yang--Mills theory with a spontaneous symmetry breaking due to a fundamental Higgs field and with the presence of the so-called ``aether term'' \cite{Carroll}, which, unlike the CFJ term, does not break the CPT symmetry. In this paper, we consider the case of three spacetime dimensions. Explicitly, our goal is to study the Yang-Mills+Higgs theory in the presence of a constant background field by means of Gribov’s confinement scenario. The constant background field is introduced within the non-Abelian aether term. As a result, we find that some regimes of the theory change if the gluon is crossing, or not, the background field. As the non-Abelian aether term in 3d was not generated before, in this paper we devote one section for its generation. Another motivation for our study is the following one: in \cite{ColKost1,ColKost2}, it was claimed that the SME can be treated as a low-energy effective limit of some fundamental theory where spontaneous LV occurs. Since the Gribov problem is a low-energy phenomenon, it is natural to expect that it can take place in LV non-Abelian theories.

First we give a review of the Gribov formalism in section \ref{reviewGr}, which is followed by section \ref{ymha} with our analysis of the Yang--Mills--Higgs--aether theory using the Gribov formalism. We end with some concluding remarks in section \ref{conc}.

\section{Review of Gribov restriction in Yang--Mills theory}
\label{reviewGr}
In this section, we present a short review of the work by Gribov presented in \cite{Gribov:1977wm}. In this paper, new way to treat the nonperturbative regime of {non-Abelian} gauge theories was proposed. This approach is based on the fact that the Faddeev--Popov procedure is not sufficient to remove all the gauge copies present in the Yang--Mills path integral. This means that an extra restriction on the gauge field in the integral is mandatory. As a consequence, a non-local term called Gribov mass arises in the system. The presence of such a term modifies the gauge propagator by removing the propagation of physical excitations. The Gribov mass vanishes in the high-energy regime of the theory and is highly relevant in the IR regime. This is Gribov's interpretation of confinement, \emph{i.e.}~the excitations of the perturbative theory are no longer present in the IR regime.
\subsection{Gribov's no-pole restriction in Yang--Mills theories}
\label{gribov} 
The Euclidean Yang--Mills path integral reads
\begin{equation}
Z_\text{YM} = \int DA \; e^{- \frac{1}{4} \int d^{d}x \; F^{a}_{\mu \nu}F^{a}_{\mu\nu}  } \;,
\label{ympathintegral}
\end{equation}
where $F^{a}_{\mu\nu}$ is the field strength tensor:
\begin{equation}
F^{a}_{\mu\nu} = \partial_{\mu}A^{a}_{\nu} - \partial_{\nu}A^{a}_{\mu} + gf^{abc}A^{b}_{\mu}A^{c}_{\nu}\;. \label{fstr}
\end{equation}
Due to the gauge redundancy present in the partition function \eqref{ympathintegral}, the Faddeev--Popov (FP) procedure is necessary. In the Landau gauge, the gauge-fixed partition function reads
\begin{equation}
Z_\text{FP} =\int DA DcD\bar{c} \; e^{-S_\text{FP}}
\;,
\label{fppathintegral}
\end{equation}
with
\begin{equation}
S_\text{FP} = 
\frac{1}{4} \int d^{d}x \; 
F^{a}_{\mu \nu}F^{a}_{\mu\nu} 
+
\int d^{d}x \;
\left(  b^{a}\partial_{\mu}A^{a}_{\mu}
+\bar{c}^{a} \partial_{\mu}D^{ab}_{\mu}c^{b}  \right)
\,.
\label{gf}
\end{equation}
The fields $({\bar c}^a, c^a)$ are the Faddeev--Popov ghosts, $b^a$ is the Lagrange multiplier implementing the Landau gauge condition:
\begin{equation}
\partial_{\mu}A_{\mu} = 0 \,,
\label{lgc}
\end{equation}
and $D^{ab}_\mu =( \delta^{ab}\partial_\mu + g f^{acb}A^{c}_{\mu})$ is the covariant derivative in the adjoint representation of $SU(N)$. In \cite{Gribov:1977wm}, it is shown that even after the FP procedure, the partition function \eqref{fppathintegral} is still plagued by the presence of some physically equivalent gauge field configurations. The proposed solution is to restrict the gauge fields to a region, namely the first Gribov region, where the FP operator ${\cal M}^{ab}$ is positive definite. This region is defined as
\begin{equation}
\Omega =  \{ A^a_{\mu}\,;\; \partial_\mu A^a_{\mu}=0 \,;\;{\cal M}^{ab} = -\partial_{\mu}(\partial_{\mu}\delta^{ab} -g f^{abc}A^{c}_{\mu})\; >0 \} \;. 
\label{gr}
\end{equation} 

As the FP operator is related to the inverse of the ghost field propagator, the extra restriction is linked to the ghost two-point function. This function can be computed as a functional of the gauge field up to one-loop order as
\begin{equation}
\mathcal{G}(k,A) = \frac{\delta^{ab}}{N^2-1}\frac{1}{k^2}\left(\delta^{ab}+\sigma^{ab}(k,A)  \right) \;,
\label{ghostformfactor1}
\end{equation}
where $\sigma(k,A)$ is called the ghost form factor. If the gauge field $A_{\mu}(x)$ has small amplitude, we have
\begin{equation}
\mathcal{G}(k,A) \approx \frac{1}{k^2}\frac{1}{\left(1-\sigma(k,A)\right)} \,.
\label{ghostformfactor}
\end{equation}
Thus, the condition to stay within the first Gribov region can also be expressed as
\begin{equation}
\sigma(k,A) < 1 \;,
\label{npcondition}
\end{equation}
This condition for the ghost form factor is known as the no-pole condition. After the FP {procedure} and constraining the path integral to a domain where the gauge field configuration satisfies the no-pole condition \eqref{ghostformfactor}, the system has no more infinitesimal gauge copies. The no-pole restriction is implemented by means of the Heaviside step function $\theta$:
\begin{align}
Z_{G} =& \int_\Omega DA DcD\bar{c} \; e^{-S_\text{FP}} \nonumber \\
=&\int DA DcD\bar{c} \;  \theta(1 - \sigma(k,A)) e^{-S_\text{FP}} \,.
\label{gribovpathintegral}
\end{align}

The ghost two-point function in the presence of an external gauge field up to first order in the quantum fields reads
\begin{equation}
\mathcal{G}(k,A)=\frac{1}{k^2}\left(1+\frac{k_\mu k_\nu}{k^2}\frac{Ng^2}{Vd(N^2-1)}\int\frac{d^dp}{(2\pi)^4}\frac{A_\mu^a(p)A_\nu^a(-p)}{(k-p)^2}\right) \,.
\label{ghostformfactor1pi}
\end{equation}
It is known it suffices to take the limit $k\to0$,\footnote{See \cite{Vandersickel:2012tz} for a detailed computation of \eqref{ghostformfactor1pi} and for further discussion.} in which case the ghost form factor reads
\begin{equation}
\sigma(0,A) = \frac{Ng^2}{dV(N^2-1)}\int\frac{d^dp}{(2\pi)^d}\frac{A_\mu^a(p)A_\mu^a(-p)}{p^2} \,.
\label{ghostformfactor2}
\end{equation}

Considering the integral representation of the Heaviside step function, the Gribov partition function becomes
\begin{equation}
Z_{G} = \int DA DcD\bar{c} \int^{\infty+i\epsilon}_{-\infty+i\epsilon}\frac{d\beta}{2\pi i\beta} \; e^{\beta(1-\sigma(0,A))}e^{-S_\text{FP}}.
\label{gribovpathintegral1}
\end{equation}
The integral over $\beta$ can be performed in the saddle-point approximation elaborated on in the next section. Finally, we can write down the Gribov action,
\begin{equation}
S_\text{G} = S_\text{FP} + \frac{\beta Ng^2}{d(N^2-1)} \int d^{d}x \; A_\mu^a(x) \left[\p^{2}\right]^{-1} A_\mu^a(x) - \beta \,.
\label{wrhglwieh}
\end{equation}

\subsection{Gap equation and gauge field propagator}
\label{glounpropagatorsec}
As can be seen from the action \eqref{wrhglwieh}, a mass parameter $\beta$ is introduced into the theory. This $\beta$ is not a free parameter of the theory. As {required by} the consistency of the model, it is determined by a gap equation.

At the tree level in perturbation theory, the partition function \eqref{gribovpathintegral1} can be written as
\begin{equation}
Z_{G} = 
\int^{\infty+i\epsilon}_{-\infty+i\epsilon} \frac{d\beta}{2\pi i\beta} \int  DA DcD\bar{c} \, \exp  \left\{-\int \frac{d^dp}{(2\pi)^d)}\left[ \frac12
A^{a}_{\mu}(p) Q^{ab}_{\mu\nu} A^{b}_{\nu}(-p) + \bar{c}^{a}(p) P^{ab} c^{b}(-p) \right] - \beta \right\} \,,
\end{equation}
with
\begin{equation}
Q_{\mu\nu}^{ab} = \delta^{ab} \left[\left(\frac{2N \beta g^2}{Vd(N^2-1)}\frac{1}{p^2}+p^2\right)\delta_{\mu\nu}+\left(\frac{1}{\alpha}-1\right)p_\mu p_\nu
\right]
\label{gluonoperator}
\end{equation}
and
\begin{equation}
P^{ab} = \delta^{ab} p^2 \,.
\end{equation}
Integrating out the fields, one ends up with
\begin{equation}
Z_{G} = \int^{\infty+i\epsilon}_{-\infty+i\epsilon} \frac{d\beta}{2 i \pi} \left[  \det  Q^{ab}_{\mu\nu} \right]^{-1/2} \left[  \det P^{ab} \right] \e^{\beta - \ln\beta} \,.
\end{equation}
Using that $(\det \mathrm M)^{-1/2}=\e^{-\frac12\tr\ln\mathrm M}$ for any matrix $\mathrm M$, one rewrites the path integral as
\begin{equation}
\e^{-V{\cal E}_{v}} = Z_{G} = \int^{\infty+i\epsilon}_{-\infty+i\epsilon}\frac{d\beta}{2 i \pi}\e^{-f(\beta)}\,,
\end{equation}
with
\begin{equation}
f(\beta) = \beta-\ln\beta-\frac{d-1}{d}(N^2-1)V\int \frac{d^dp}{(2\pi)^d}\ln\left(p^2+\frac{\beta Ng^2}{N^2-1}\frac{2}{dV}\frac{1}{p^2}\right)
\,.
\label{fbeta}
\end{equation}
As mentioned before, by means of the saddle-point approximation we have
\begin{equation}
Z_{G} \approx e^{-f(\beta^{\ast})},
\end{equation}
where $\beta^{\ast}$ is that value of $\beta$ which satisfies the saddle-point equation
\begin{equation}
\frac{\p {\cal E}_{v}}{\p \beta} \Bigg \vert_{\beta = \beta^{\ast}} = 0 \,,
\end{equation}
leading us to the so-called gap equation\footnote{In the thermodynamic limit, the term $\ln\beta$ can be disregarded, since $\beta$ is proportional to the volume $V$.}
\begin{equation}
1 = \frac{d-1}{d}Ng^2 \int \frac{d^dp}{(2\pi)^d)}\frac{1}{p^4+\gamma^{\ast \, 4}} \,.
\label{gapequation}
\end{equation}
In order to simplify the notation, we defined $\gamma^4=\frac{2g^2\beta N}{dV(N^2-1)}$. As mentioned before, the gap equation \eqref{gapequation} can be seen as a self-consistency condition of the model, \emph{i.e.}~the Gribov mass parameter is determined by \eqref{gapequation}.

From \eqref{gluonoperator}, it can be seen that the gluon propagator is influenced by the Gribov parameter $\gamma^4$. The gluon propagator in the Landau limit $\Delta \to 0$ reads
\begin{align}
\langle A_\mu^a(k)A_\nu^b(-k)\rangle =& \delta^{ab}\frac{k^2}{k^4+\gamma^{\ast \, 4}} \left(\delta_{\mu\nu}-\frac{k_\mu k_\nu}{k^2}\right) \nonumber \\
= & \delta^{ab} \frac{1}{2}\left(\frac{1}{k^2+i\gamma^{\ast \, 2}}+\frac{1}{k^2-i\gamma^{\ast \, 2}}\right) \left(\delta_{\mu\nu}-\frac{k_\mu k_\nu}{k^2}\right) \,.
\label{gribovpropagator}
\end{align}
Thus, from \eqref{gribovpropagator}, it is clear that due to Gribov's restriction, the gluon propagator
displays complex conjugate poles. This prevents us from assigning an asymptotic single-particle interpretation to the gluon propagator (its K\"all\'en-Lehmann representation is not always positive \cite{Sorella:2010it}). As was mentioned before, Gribov interpreted this as confinement, \emph{i.e.}~the excitations of the perturbative theory are physically absent in the IR regime.

\section{The $3D$ Yang--Mills--aether theory with Higgs fields}
\label{ymha}
In this section, in order to describe a mechanism through which the desired non-Abelian aether term arises in $3D$, we describe its perturbative generation, and, by means of the approach described in the previous section,  investigate the different regimes of the $3D$ Yang--Mills--Higgs--aether theory.

\subsection{Perturbative generation of the aether term in $3D$}
Various Lorentz-breaking terms, including the aether term, either Abelian or non-Abelian ones, can be generated perturbatively as one-loop corrections in some theory involving a (non-Abelian) gauge field coupled to spinors with inclusion of Lorentz-breaking parameters. This methodology was proposed already in \cite{ColKost2} and applied to the Abelian aether term in various space-time dimensions from 3 to 5 in \cite{Gomes:2009ch,aether2}, and to its four-dimensional non-Abelian analogue in \cite{ournAb}.
Now, let us discuss the three-dimensional non-Abelian aether term. We start with the extended spinor QED action which can be treated as a natural $3D$ analogue of the action proposed in \cite{aether2}:
\begin{equation}
S=\int d^3x\bar{\psi}^i\left(i\gamma^{\mu}(\partial_{\mu}\delta^{ij}-ieA_{\mu}^a(T^a)^{ij}+b_{\mu}\delta^{ij})+g\epsilon^{\mu\nu\lambda}b_{\mu}F_{\nu\lambda}^a(T^a)^{ij}-m\delta^{ij}\right)\psi^j,
\end{equation}
where $(T^a)^{ij}$ are the generators of the corresponding Lie algebra, and both the gauge field $A_{\mu}=A_{\mu}^aT^a$ and the non-Abelian stress tensor $F_{\mu\nu}=F_{\mu\nu}^aT^a$ are Lie-algebra valued. The vertex involving $A_{\mu}$ is further referred to as the minimal one, and the one involving the $F_{\mu\nu}$ as the non-minimal one. We note that, unlike in the four-dimensional case, there is no chirality in three dimensions, and the analogue of the $\gamma_5$ matrix given by $\gamma_0\gamma_1\gamma_2$ is proportional to the unit matrix. It must be emphasized that the presence of the gauge covariant derivative $D_{\mu}^{ij}=\partial_{\mu}\delta^{ij}-ieA_{\mu}^a(T^a)^{ij}$ is necessary for the full-fledged gauge invariance of the action while otherwise,  in the absence of the minimal coupling, only restricted gauge invariance, with constant gauge parameters without dependence on the space-time coordinates, is possible. 

Completely analogously to \cite{aether2}, there could be three contributions to the aether term -- the one formed by two minimal vertices, the mixed one, and the one formed by two non-minimal vertices. However, straightforward calculations show that the purely minimal contribution vanishes. This fact can be justified as follows: the integral over momenta in the corresponding Feynman diagram is the same for the Abelian and the non-Abelian case, and the Abelian minimal aether-like contribution (proportional to $e^2$) is zero since in this case the Lorentz-breaking vector $b_{\mu}$ is ruled out by the simple gauge transformation $A_{\mu}\to A_{\mu}-b_{\mu}$. Hence the minimal contribution of second order in $b_{\mu}$ is zero independently on the gauge group. The mixed aether contribution (proportional to $eg$) also can easily be shown to vanish in the Abelian case (indeed, the only relevant term, after the above-mentioned gauge transformation,  turns out to be of first order in $b_{\mu}$), hence, by gauge symmetry reasons, the non-Abelian generalization of this term will also absent (indeed, the Abelian contribution is the quadratic part of the non-Abelian one).

Therefore, the only nontrivial aether-like contribution is the purely non-minimal one. In this case we can straightforwardly apply the results obtained in \cite{Gomes:2009ch}, with the only difference a factor $\kappa$ arising from the definition of the trace ${\rm tr}(T^aT^b)=\kappa\delta^{ab}$ and coming from the product of two generators in two vertices, and write down the desired aether term
\begin{equation}
S_{aether}=\frac{4|m|g^2\kappa}{2\pi}b^{\mu}F_{\mu\nu}^ab_{\lambda}F^{\lambda\nu a}.
\end{equation}
So, we explicitly demonstrated how the $3D$ non-Abelian aether term arises. It is clear that in four dimensions, the non-Abelian aether contributions will be generated for all three cases, not only the minimal one studied in \cite{ournAb}. However, the contributions involving either one or two non-minimal vertices will be ambiguous in full analogy with \cite{aether2}.

\subsection{Gauge propagator}
The Yang--Mills--Higgs--aether Euclidean action before gauge fixing and implementing the Gribov formalism reads
\begin{equation}
S=\int d^3x \left(\frac{1}{4}\left(F_{\mu\nu}^a\right)^2+\frac{\alpha}{2}a_\mu F_{\mu\nu}^a a_\delta F_{\delta\nu}^a\right)+(D_\mu^{ij}\Phi^j)^\dagger(D_\mu^{ik}\Phi^k)+\frac\lambda2\left(\Phi^\dagger\Phi-\nu^2\right)^2 \;.
\label{ymlvaction}
\end{equation}
To this action, we add a Landau gauge fixing term and the contributions coming from the Gribov formalism.

We can write the quadratic terms as
\begin{equation}
S_\text{quad} = \int \frac{d^3k}{(2\pi)^4}\left(\frac{1}{2}{A}_\mu^a(k)Q_{\mu\nu}^{ab} {A}_\nu^b(-k)\right) \;,
\end{equation}
where we introduced the inverse propagator operator 
\begin{eqnarray}
Q_{\mu\nu}^{ab}&=&\delta^{ab}\left[\left(k^2+\frac{\gamma^4}{k^2}+\frac{g^2\nu^2}{2}\right)\delta_{\mu\nu}+\left(\frac{1}{\Delta}-1\right)k_\mu k_\nu+\right.\nonumber\\&+& \left.\alpha\left((a\cdot k)^2 \delta_{\mu\nu}-(a\cdot k) a_\nu k_\mu - a_\mu k_\nu (a\cdot k)+k^2 a_\mu a_\nu\right)\right] \;,
\label{opq}
\end{eqnarray}
where we still have $\gamma^4=\frac{\beta Ng^2}{2V(N^2-1)}$ the Gribov parameter and $\Delta$ the gauge fixing parameter, which must be put to zero for the Landau gauge.

In this $\Delta\to0$ limit, the tree-level gauge propagator reads
\begin{equation}
\langle A_\mu^a(k)A_\nu^b(k)\rangle = \delta^{ab}F_1(k)\left[\left(\delta_{\mu\nu}-\frac{k_\mu k_\nu}{k^2}\right)- F_2(k) \left((a\cdot k)k_\mu-k^2a_\mu\right)\left((a\cdot k)k_\nu-k^2a_\nu\right)\right] \;,
\end{equation}
where
\begin{subequations} \begin{gather}
F_1(k)=\frac{k^2}{k^4+{\gamma^4}+\frac{g^2\nu^2}2 k^2+\alpha(a.k)^2k^2} \;, \\
F_2(k) = \frac\alpha{(1+\alpha a^2)k^4+\gamma^4+\frac{g^2\nu^2}2 k^2} \;.
\end{gather} \label{fs} \end{subequations}
The poles of $F_1(k)$ are found at $k^2$ equal to minus
\begin{equation}
m^2_\pm = -\frac{g^2\nu^2}{4\xi(\theta)}\pm\frac{1}{4\xi(\theta)}\sqrt{\left({g^4\nu^4}-16\xi(\theta){\gamma^4}\right)}
\label{solutionpoles}
\end{equation}
where we defined $\theta$ as the angle between $k_\mu$ and $a_\mu$, and $\xi(\theta)=1+\alpha a^2\cos^2\theta$. The poles of $F_2(k)$ can be found by putting $\theta=0$.

\subsection{The gap equation}
\label{gapequationaether}
Repeating the steps that led to \eqref{gapequation}, one finds that the gap equation in our case is given by
\begin{equation}
	d = \frac{Ng^2}{V(N^2-1)} \Tr (Q_{\mu\nu}^{ab})^{-1} \;,
\end{equation}
where $Q_{\mu\nu}^{ab}$ was defined in \eqref{opq}.

To compute the trace, we use a basis in which $Q_{\mu\nu}^{ab}$ is diagonal. For any vector $v_\mu$ orthogonal both {\bf to} $k_\mu$ and $a_\mu$, we have
\begin{equation}
	Q_{\mu\nu}^{ab} v_\nu = \delta^{ab} \left( k^2+\frac{\gamma^4}{k^2} + \frac{g^2\nu^2}2 + \alpha (a\cdot k)^2\right) v_\mu \;,
\end{equation}
which give{\color{red}s} us the first $d-2$ eigenvalues. Besides that, for $k_\mu$ we have:
\begin{equation}
	Q_{\mu\nu}^{ab} k_\nu = \delta^{ab} \left( \frac{\gamma^4}{k^2}+\frac{g^2\nu^2}2 + \frac1\Delta k^2\right) k_\mu \;{\color{red}.}
\end{equation}
This yields one more eigenvalue. In order to find the last eigenvalue, we consider a vector in the $(a_\mu,k_\mu)$  plane but orthogonal to $k_\mu$:
\begin{equation}
	Q_{\mu\nu}^{ab} \left(a_\nu-\frac{a\cdot k}{k^2} k_\nu\right) = \delta^{ab} \left( k^2+\frac{\gamma^4}{k^2}+\frac{g^2\nu^2}{2} + \alpha k^2a^2\right) \left(a_\mu-\frac{a\cdot k}{k^2} k_\mu\right) \;.
\end{equation}
To conclude, we have:
\begin{multline}
	\Tr (Q_{\mu\nu}^{ab})^{-1} = V(N^2-1) \left[ (d-2) \int\frac{d^dk}{(2\pi)^d} \frac1{k^2+\frac{\gamma^4}{k^2} + \frac{g^2\nu^2}{2}+ \alpha (a\cdot k)^2} + \int\frac{d^dk}{(2\pi)^d} \frac1{\frac{\gamma^4}{k^2}+ \frac{g^2\nu^2}{2} + \frac1\Delta k^2} \right. \\ \left. + \int\frac{d^dk}{(2\pi)^d} \frac1{k^2+\frac{\gamma^4}{k^2}+\frac{g^2\nu^2}{2} + \alpha a^2k^2} \right] \;.
\end{multline}
The second term is zero in the limit $\Delta\to0$. For $d=3$, the integrals are furthermore finite as is usual in odd-dimensional space-times, making the dimensional regularization unnecessary. This leaves us with the following gap equation:
\begin{equation}
	3 = Ng^2 \left( \int \frac{d^3k}{(2\pi)^3} \frac1{\xi(\theta)k^4+\frac{g^2\nu^2}2 k^2+\gamma^4} + \int \frac{d^3k}{(2\pi)^3}\frac1{\zeta k^4+\frac{g^2\nu^2}2 k^2+\gamma^4} \right) \;, \label{finalgap}
\end{equation}
where we put $\zeta=1+\alpha a^2$ and we still have $\xi(\theta)=1+\alpha a^2\cos^2\theta$.

The integrals on the r.h.s. are finite, such that we can argue using normal integration rules:
\begin{itemize}
\item Taking the derivative of the r.h.s. with respect to $\gamma^4$ yields integrals of minus a square, which is negative. This means that the r.h.s. of the gap equation decreases with $\gamma$.
\item After changing variables $k_\mu \to \gamma q_\mu$, it is possible to expand the integrals with respect to large $\gamma$, yielding results behaving as $1/\gamma$. This means that the r.h.s. goes to zero for large $\gamma$.
\end{itemize}
As a result, the r.h.s. is a {monotonously} decreasing function of $\gamma$, with limiting value zero. The gap equation can therefore only have a solution for real (positive) $\gamma$ if and only if the r.h.s. of the gap equation evaluated at $\gamma=0$ is more than 3:
\begin{equation}
	3 < Ng^2 \left( \int \frac{d^3k}{(2\pi)^3} \frac1{\xi(\theta)k^4+\frac{g^2\nu^2}2 k^2} + \int \frac{d^3k}{(2\pi)^3}\frac1{\zeta k^4+\frac{g^2\nu^2}2 k^2} \right) \;.
\end{equation}
If this inequality is not satisfied, there can be no physical solutions, but one checks easily that the no-pole condition \eqref{npcondition} is automatically satisfied in that case, and the Gribov formalism is therefore not needed.

The integrand of the second integral can be rewritten using the Feynman parametrization:
\begin{equation}
	\frac1\zeta \frac1{k^2(k^2+\frac{g^2\nu^2}{2\zeta})} = \frac1\zeta \int_0^1 \frac{dx}{(k^2+x\frac{g^2\nu^2}{2\zeta})^2}
\end{equation}
Integrating over $k_\mu$ and then over $x$ immediately yields
\begin{equation}
	\frac1{2\pi g\nu\sqrt{2\zeta}} \;.
\end{equation}
For the first integral, use spherical coordinates and that
\begin{subequations} \begin{gather}
	\int \frac{dx}{\alpha x^2+\beta} = \frac1{\sqrt{\alpha\beta}} \arctan(x\sqrt{\tfrac\alpha\beta}) + C \;, \\
	\int \frac{dx}{\sqrt{\alpha x^2+1}} = \frac1{\sqrt\alpha} \operatorname{arsinh}(x\sqrt\alpha) + C \;,
\end{gather} \end{subequations}
to find
\begin{equation}
	\frac1{2\pi g\nu a\sqrt{2\alpha}} \operatorname{arsinh}(a\sqrt{\alpha}) \;.
\end{equation}

\begin{figure}[h]
\begin{center}
\includegraphics[width=.5\textwidth]{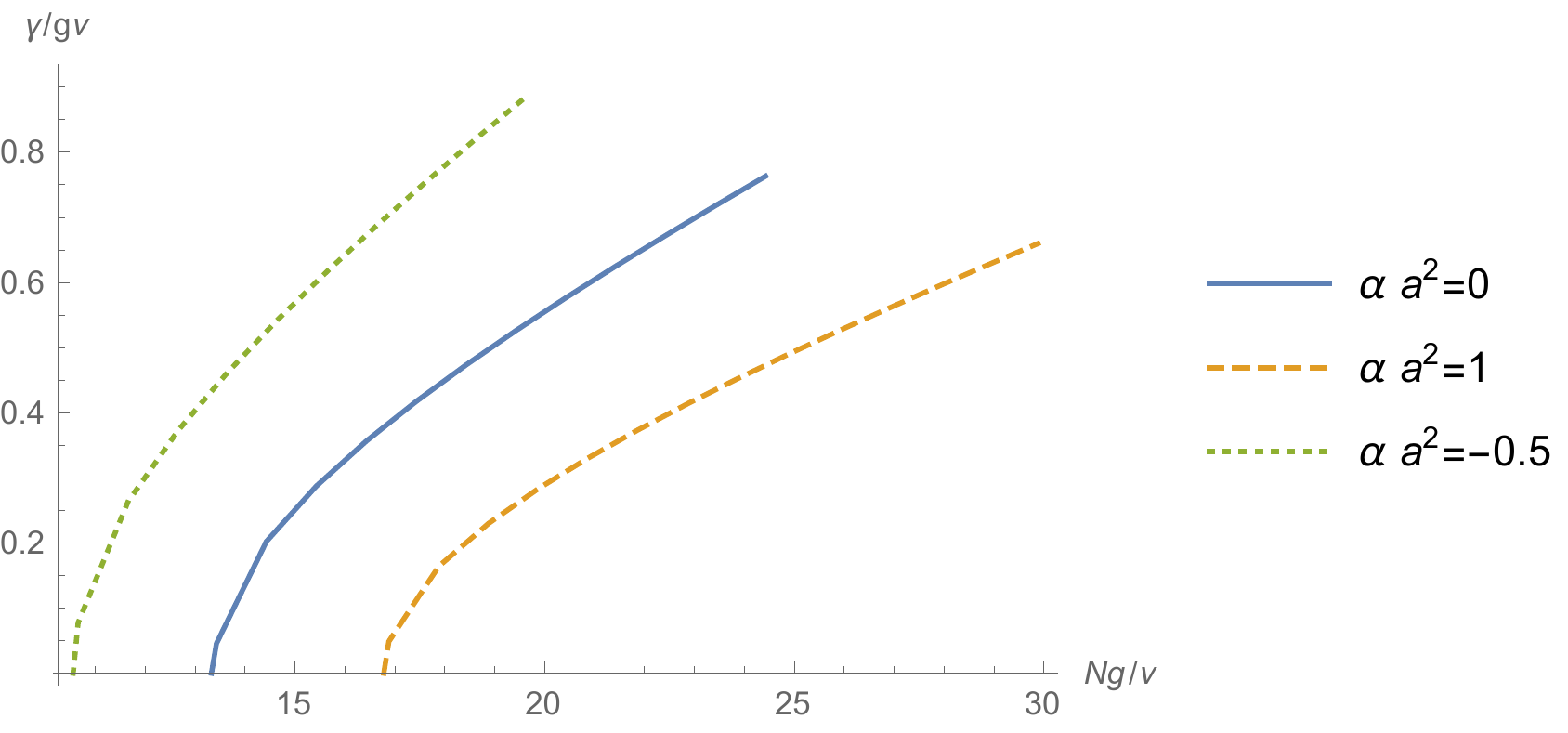}
\end{center}
\caption{Values of the dimensionless Gribov parameter $\gamma/g\nu$ as function of $Ng/\nu$ from a numerical solution of \eqref{finalgap} for select values of the Lorentz breaking $\alpha a^2$.} \label{gamma_vs_g}
\end{figure}

\begin{figure}[h]
\begin{center}
\includegraphics[width=.5\textwidth]{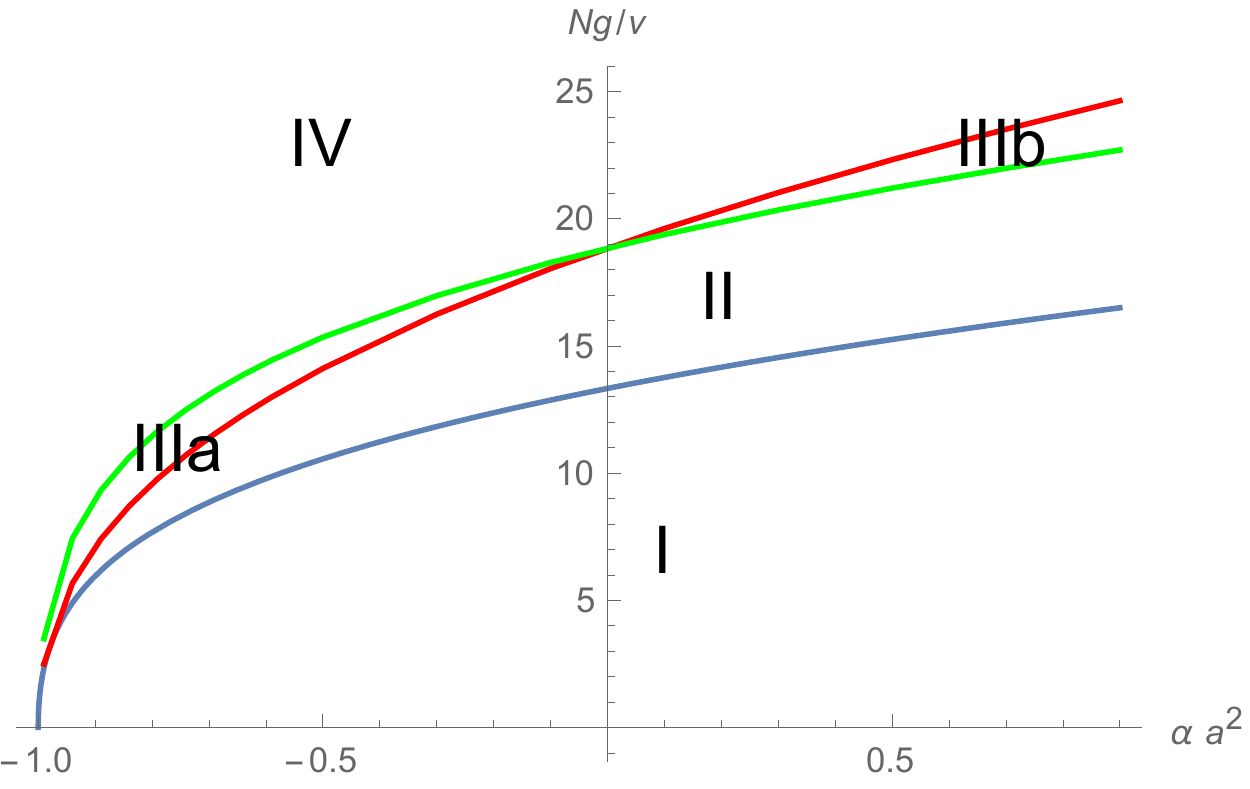}
\end{center}
\caption{Boundaries between different behaviors of the poles of the gluon propagator \eqref{solutionpoles}. See section \ref{analysis} for an in-depth discussion.}
\label{poles}
\end{figure}

The condition for a solution to the gap equation is therefore
\begin{equation}
	3 < \frac{Ng}{2\pi\nu\sqrt2} \left( \frac{\operatorname{arsinh}(a\sqrt{\alpha})}{a\sqrt{\alpha}} + \frac1{\sqrt{1+\alpha a^2}} \right),
\end{equation}
or, for a small Lorentz-breaking term, which is the physical case,
\begin{equation}
	g > \frac{3\pi\nu\sqrt2}N \left(1+\frac{\alpha a^2}3+\cdots\right) \;.
\end{equation}
This is the blue line separating regimes I and II in \figurename\ \ref{poles}: there is no nontrivial Gribov parameter in the weak-coupling strong-Higgs regime I, while the Gribov parameter is nonzero (regimes II to IV) for strong coupling or weak Higgs term. Some numerically determined values of the Gribov parameter are plotted in \figurename\ \ref{gamma_vs_g}. One notices that a positive value for $\alpha a^2$ pulls the Gribov parameter down, while a negative aether term pushes it up.

\subsection{Regimes}
\label{analysis}

From the above, we conclude there are at least two regimes: one with and one without Gribov parameter. The situation becomes more complex when looking at the behavior of the gluon propagator. In the absence of a nonzero Gribov parameter, the gluon simply behaves like a massive but Lorentz-broken Higgs type gluon.

Once the Gribov parameter is nonzero, this begins to change. At not very high values of the Gribov parameter, however, the gluon propagator still has real, massive poles like in the pure Higgs--aether case. This is what happens in regime II of \figurename\ \ref{poles}. At some point, however, the discriminant of the denominators of \eqref{fs} changes sign, at least for some values of $\theta$. The green line in  \figurename\ \ref{poles} indicates a change in sign for $\theta=0$ (and thus also for the denominator of $F_2(k)$), while the red line indicates a change in sign for $\theta=\pi$. The sign changes for other values of $\theta$ happen between these two lines.

As a result we have the regimes labeled IIIa and IIIb, where the gluon has real or complex poles depending on the direction of propagation relative to the aether field. In the strong-coupling weak-Higgs regime IV, all poles have nonzero imaginary parts, and no physical gluons can propagate.

One sees that all the transition lines in \figurename\ \ref{poles} increase with $\alpha a^2$. As a result, introducing a positive aether term has a qualitative effect comparable to lowering the coupling strength or increasing the strength of the Higgs background, while a negative aether term is similar to stronger coupling or weaker Higgs.

\section{Conclusion}
\label{conc}
In this work we presented a first study of non-Abelian Yang--Mills--Higgs with an additive Lorentz breaking aether term in 3 spacetime dimensions. To justify the presence of the non-Abelian aether term, we performed its perturbative generation, demonstrating that it arises as a one-loop correction. As a by-product, we argued that the non-Abelian aether-like term in $4D$ receives contribution from non-minimal couplings, too. We used the Gribov--Zwanziger formalism to fix the gauge in Landau gauge without infinitesimal gauge copies, which gives insight in the nonperturbative dynamics of the theory.

We found that a positive aether parameter $\alpha a^2$ reduces the value of the Gribov parameter and (if sufficiently large) can turn the theory from showing a nonpertubative behavior to a perturbative one. A negative value of $\alpha a^2$ has the opposite effect.

Compared to the 3D $SU(N)$ Yang--Mills--Higgs case, we find one extra intermediate regime: in between the regime with real poles and the one with complex poles in the gluon propagator, there is an additional regime where the reality of the gluon propagator poles depends on the direction of propagation. The different regimes are depicted in \figurename\ \ref{poles}.

A next step in this line of research would be to consider the theory in an appropriate temporal background which allows to access the vacuum expectation value of the Polyakov loop \cite{Marhauser:2008fz,Braun:2007bx,Reinhardt:2013iia,Reinosa:2014ooa}, which would tell us whether the theory is effectively confined in the regime where the gluon propagator has complex poles and deconfined otherwise.

\section*{Acknowledgments}
The work by A. Yu. P. has been partially supported by CNPq, project 301562/2019-9.

\end{document}